\documentclass[12pt]{spieman}  
\usepackage[utf8]{inputenc}   
\usepackage{soul} 
\usepackage{aas_macros}
\usepackage{graphicx}
\usepackage{paralist}
\usepackage{tabularx}

\renewcommand{\hl}{} 

\title{The SOXS Spectrograph Instrument Control Software}

\author[*,a]{Davide Ricci}
\author[a]{Bernardo Salasnich}
\author[a]{Andrea Baruffolo}

\author[l]{Jani Achrén}
\author[b]{Matteo Aliverti}
\author[d]{José A. Araiza-Durán}
\author[n]{Iair Arcavi}
\author[b]{Laura Asquini}
\author[a]{Federico Battaini}
\author[g]{Sagi Ben-Ami}
\author[g]{Alex Bichkovsky}
\author[v]{Anna Brucalassi}
\author[g]{Rachel Bruch}
\author[b]{Lorenzo Cabona}
\author[b]{Sergio Campana}
\author[c]{Giulio Capasso}
\author[a]{Enrico Cappellaro}
\author[a]{Riccardo Claudi}
\author[c]{Mirko Colapietro}
\author[e]{Rosario Cosentino}
\author[f]{Francesco D'Alessio}
\author[b]{Paolo D'Avanzo}
\author[c]{Sergio D'Orsi}
\author[c]{Massimo Della Valle}
\author[k]{Rosario Di Benedetto}
\author[a]{Simone Di Filippo}
\author[g]{Avishay Gal-Yam}
\author[b]{Matteo Genoni}
\author[e]{Marcos Hernandez Díaz}
\author[g]{Ofir Hershko}
\author[j,q]{Jari Kotilainen}
\author[j,q]{Hanindyo Kuncarayakti}
\author[b]{Marco Landoni}
\author[r]{Gianluca Li Causi}
\author[c]{Laurent Marty}
\author[q]{Seppo Mattila}
\author[k]{Matteo Munari}
\author[b]{Luca Oggioni}
\author[e]{Hector Pérez Ventura}
\author[b]{Giorgio Pariani}
\author[m]{Giuliano Pignata}
\author[a]{Kalyan~Kumar Radhakrishnan~Santhakumari}
\author[s]{Stephen Smartt}
\author[g]{Michael Rappaport}
\author[b]{Marco Riva}
\author[h]{Adam Rubin}
\author[c]{Salvatore Savarese}
\author[c]{Pietro Schipani}
\author[w,k]{Salvatore Scuderi}
\author[u]{Maximilian Stritzinger}
\author[f]{Fabrizio Vitali}
\author[s]{David Young}
\author[k]{Ricardo Zanmar Sanchez}
\author[h]{Gerard Zins}

\affil[a]{INAF -- Osservatorio Astronomico di Padova, Vicolo dell’Osservatorio 5, I-35122, Padua, Italy }
\affil[b]{INAF -- Osservatorio Astronomico di Brera, Via Bianchi 46, I-23807, Merate, Italy }
\affil[c]{INAF -- Osservatorio Astronomico di Capodimonte, Salita Moiariello 16, I-80131, Naples, Italy }
\affil[d]{Instituto de Astronom\'ia - Universidad Nacional Aut\'onoma de M\'exico , Km 103 Carretera Tijuana-Ensenada, Ensenada, 22860, Mexico}
\affil[e]{FGG-INAF, TNG, Rambla J.A. Fernández Pérez 7, E-38712 Breña Baja (TF), Spain }
\affil[f]{INAF -- Osservatorio Astronomico di Roma, Via Frascati 33, I-00078 M. Porzio Catone, Italy }
\affil[g]{Weizmann Institute of Science, Herzl St 234, Rehovot, 7610001, Israel }
\affil[h]{ESO, Karl Schwarzschild Strasse 2, D-85748, Garching bei München, Germany }
\affil[i]{Max-Planck-Institut für Extraterrestrische Physik, Giessenbachstr. 1, D-85748 Garching, Germany }
\affil[j]{Finnish Centre for Astronomy with ESO (FINCA), FI-20014 University of Turku, Finland}
\affil[k]{INAF -- Osservatorio Astrofisico di Catania, Via S. Sofia 78 30, I-95123 Catania, Italy }
\affil[l]{Incident Angle Oy, Capsiankatu 4 A 29, FI-20320 Turku, Finland }
\affil[m]{Instituto de Alta Investigaci\'on, Universidad de Tarapac\'a, Arica, Casilla 7D, Chile}
\affil[n]{The School of Physics and Astronomy, Tel Aviv University, Tel Aviv 69978, Israel}
\affil[o]{Dark Cosmology Centre, Juliane Maries Vej 30, DK-2100 Copenhagen, Denmark }
\affil[p]{Aboa Space Research Oy, Tierankatu 4B, FI-20520 Turku, Finland}
\affil[q]{Tuorla Observatory, Dept. of Physics and Astronomy, FI-20014 University of Turku, Finland }
\affil[r]{INAF - Istituto di Astrofisica e Planetologia Spaziali,  Via del Fosso del Cavaliere 100, I-00133, Rome, Italy}
\affil[s]{Astrophysics Research Centre, Queen's University Belfast, Belfast, BT7 1NN, UK }
\affil[u]{Aarhus University, Ny Munkegade 120, D-8000 Aarhus, Denmark }
\affil[v]{INAF-Osservatorio Astrofisico di Arcetri,  Largo Enrico Fermi 5, I-50125, Florence, Italy}
\affil[w]{INAF -- Istituto di Astrofisica Spaziale e Fisica Cosmica, Milan, Via Corti 12, I-20133 Milano, Italy}

\begin{document}
\maketitle

\begin{abstract}
  
  SOXS (Son Of X-Shooter) is a new spectrograph for the European
  Southern Observatory (ESO), recently installed at the New Technology
  Telescope (NTT) at the La Silla Observatory, Chile.
  
  The main instrument goal consists in the characterization of
  transient sources, based on alerts.  It covers from (partially)
  ultra-violet to visible and near-infrared bands, with a spectral
  resolution of $R\sim 4500$, using two separate, wavelength-optimized
  spectrographs.  A scientific grade visible camera, primarily
  intended for target acquisition, also provides a ``light imaging''
  mode.
  
  In this paper, we present the design of the SOXS Instrument Control
  Software, which is in charge of controlling all motors, calibration
  lamps and detectors, monitoring sensors and components' status,
  coordinating the execution of exposures, and implementing all
  observation, calibration and maintenance procedures.
  
  Given the extensive experience of the SOXS consortium in the
  development of instruments for the ESO Very Large Telescope (VLT),
  we decided to base the design of the Control System on the same
  standards, both for hardware and software control.
  We illustrate the control network, the instrument functions and
  detectors to be controlled, the overall design of SOXS Instrument
  Software (INS) and its main components.
  Then, we provide details about the control software for the most
  SOXS-specific components, and peculiar features: the piezoelectric
  tip-tilt corrector used for active compensation of mechanical
  flexures of the instrument; the cryogenic piezoelectric slit
  exchanger for the NIR spectrograph; the co-rotator monitoring
  system; and the control of the Commercial-Off-the-Shelf (COTS)-based
  imaging camera.
    
\end{abstract}


\section{Introduction}
\label{sec:intro}

SOXS, which stands for ``Son Of X-Shooter'', is a new
instrument\cite{
  soxs-campana,
  2022SPIE12184E..0OS,
  2020SPIE11447E..09S,
  2016SPIE.9908E..41S}
developed for the European
Southern Observatory's (ESO) New Technology Telescope (NTT) at the La
Silla Observatory, Chile. It is inspired by the X-Shooter
spectrograph\cite{2011A&A...536A.105V} at the Very Large Telescope
(VLT).
This new, transient-oriented facility\cite{
  2024SPIE13096E..1TS,
  2019frap.confE..78C,
  2019vltt.confE..28R,
  2019EPSC...13.1955C,
  2019eeu..confE..12C,
  2024SPIE13096E..73R}
is mainly dedicated to several follow-up
programs\cite{2018SPIE10702E..0FS} for the characterization of sources
based on alerts.  These alerts come from ``traditional'' telescope
surveys ranging from high-energy to radio astronomy, including
neutrinos and gravitational wave experiments
\cite{2018MNRAS.474..411B}.
The wavelength coverage of the instrument span from a part of the
ultra-violet (UV) spectrum range, to the visible and the near-infrared
bands\cite{2018SPIE10702E..27Z} (overall $350$ -- $2000 \rm nm$). The
instrument has a spectral resolution of $R\sim 4500$.  It is installed
at one of the two Nasmyth foci of NTT and it features two
spectrographs: the first is optimized for the near-infrared
wavelengths, while the second is optimized for visible wavelengths and
the concerned ultra-violet range. Hereafter, we will refer to this
part of the instrument as ``visible spectrograph''.  An imaging camera
also provides a scientific ``light imaging'' mode, as well as support
for target acquisition.

\begin{figure}[t]
  \centering
  \includegraphics[width=\textwidth]{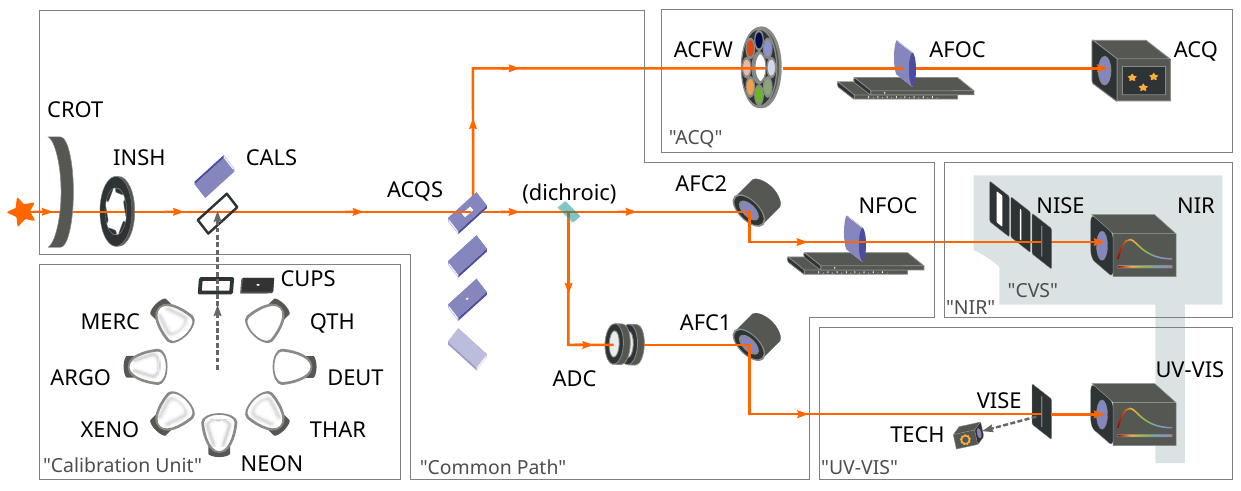}
  \caption[SOXS schematic view]
  { \label{fig:opto} Schematic view of the SOXS instrument and its
    subsystems. Gray, quoted labels refer to subsystem names. The
    filled box includes the parts under cryogenics.  Details of the
    components are described in Sect.~\ref{sec:intro}. In particular,
    for clarity, only three of the NISE slit positions and one the
    VISE slit positions are displayed in the picture.}
\end{figure}

SOXS hardware is composed of several sub-systems\cite{
  2020SPIE11447E..6OA,
  2018SPIE10702E..31A} (see boxes in Fig.~\ref{fig:opto}):
\begin{inparaitem}[\null]
\item the Common Path (CP)\cite{
    2022SPIE12184E..82R,
    2020SPIE11447E..6PB,
    2018SPIE10702E..3TC};
\item the Calibration Unit (CU)\cite{
    2020SPIE11447E..66K}.
\item the Acquisition Camera (ACQ)\cite{
    2024SPIE13096E..72A,
    2022SPIE12184E..83A,
    2020SPIE11447E..5VB,
    2018SPIE10702E..2MB};
\item the Near-Infrared spectrograph (NIR)\cite{
    2024SPIE13096E..2VV,
    2024SPIE13096E..2TG,
    2022SPIE12184E..7ZV,
    2020SPIE11447E..5NV,
    2018SPIE10702E..28V};
\item the Visible spectrograph (UV-VIS)\cite{
    2024SPIE13096E..2UC,
    2022SPIE12184E..5IC,
    2020SPIE11447E..6CC,
    2020SPIE11447E..5LR,
    2018SPIE10702E..2ZR,
    2018SPIE10702E..2JC};
\end{inparaitem}
and the Cryo-Vacuum System (CVS)\cite{ 2022SPIE12188E..44S}, that is
considered as an auxiliary system.
Here we briefly describe each one, focusing on the components that
are controlled and monitored by the Instrument Control Software, and
we summarize them in Table~\ref{tab:devices}.

\begin{description}
\item[Common Path (CP):] \hl{responsible for receiving the $f/11$
    light beam from NTT delivering a $f/6.5$ beam to NIR and VIS
    spectrographs. In doing so, the CP selects the wavelength range
    for the spectrographs using a dichroic and changes the focal
    ratio.  A number of motor devices are present in the CP:}
  
  \begin{itemize}
  \item 
    the entrance instrument shutter (INSH) is used to let
    the telescope beam in or make the instrument light-tight when
    performing calibrations using internal light sources;
  \item 
    a ``Calibration Slide'' linear motor (CALS) \hl{allows the
      selection of the instrument input source, either the light
      coming from the target field on sky or the lamps of the
      Calibration Unit;}
  \item 
    \hl{then, an ``Acquisition Slide'' linear motor (ACQS) directs the
      full beam either to the Acquisition Camera for imaging
      observations or splits it such that the central portion is
      transmitted to the spectrographs while the peripheral light is
      sent to the Acquisition Camera}\cite{2020SPIE11447E..7CC}. Two
    additional positions are used for calibration and maintenance;
  \item 
    the light directed to the spectrographs \hl{is then split by a
    dichroic into near-infrared and visible components. Each of the
    two light paths reach a separate ``Active Flexure Compensators''
    (AFC1 and AFC2), piezo-electric Tip-Tilt Mirrors responsible for
    compensating mechanical flexures due to the changing of the gravity
    vector during observations}\cite{2020SPIE11447E..5FZ};
  \item 
    a ``Near-infrared Focuser'' linear motor (NFOC), mounting an
    optical doublet, provides adjustment of the focus for the
    near-infrared light beam before reaching the corresponding
    spectrograph;
  \item
    the visible light beam passes through an ``Atmospheric Dispersion
    Corrector'' composed of two counter-rotating prisms (ADC) before
    reaching the corresponding spectrograph\cite{
      2022SPIE12184E..80B};
  \item
    the SOXS instrument is attached to one of \hl{the NTT rotator
      adapter}. A co-rotator motor ensures that all instrument cables
    are properly counter-rotated before reaching the electronic
    cabinets.
    
  \end{itemize}

\begin{table}[p]
  \begin{tabularx}{\textwidth}{clX}
\hline
\textbf{Name} &          \textbf{Device}          & \textbf{Options}                                            \\
\hline
    INSH      &        Instrument Shutter         & 2 positions: Open, Close.                                    \\
    CALS      &         Calibration Slide         & 2 positions: Calibration Unit, Science.                      \\
    CUPS      &  Calibration Unit Pinhole Slide   & 2 positions: Free, Pinhole.                                  \\
    ACQS      &     Acquisition Camera Slide      & 4 positions: Imaging, Spectroscopy, Pinhole, Pellicle.       \\
   AFC1-2     & UV-VIS and NIR channels Actuator Piezos  & 3 modes: AUTO (continuous compensation of mechanical flexures based on look-up table), STAT (stopped on a position), REF (stopped on a reference position).        \\
    NFOC      &          NIR Focus Slide          & Continuous.                                                  \\
   ADC1-2     &    Dispersion Corrector Prisms    & 3 modes: AUTO (continuous compensation of atmospheric dispersion based on telescope position), OFF (stopped), REF (stopped in a reference position).      \\
    AFOC      &  Acquisition Camera Focus Slide   & Continuous.                                                  \\
    ACFW      &  Acquisition Camera Filter Wheel  & 8 positions: SDSS $u$, $g$, $r$, $i$, $z$; Vimos $V$, Vimos $Y$, Free.            \\
    NISE      &   Near Infrared Slit Exchanger    & 6 positions: $0.5^{\prime\prime}$, $1.0^{\prime\prime}$, $1.5^{\prime\prime}$, $5.0^{\prime\prime}$ slits; pinhole, Multiple Pinhole, and BLANK (totally covered) position.              \\
    VISE      &    UV-VIS Slit Exchanger    & 7 positions: $0.5^{\prime\prime}$, $1.0^{\prime\prime}$, $1.5^{\prime\prime}$, $5.0^{\prime\prime}$ slits; pinhole, Multiple Pinhole, Slit Viewer. \\
\hline
    MERC      &            Hg Arc Lamp            & 2 states: On, Off.                                           \\
    ARGO      &            Ar Arc Lamp            & 2 states: On, Off.                                           \\
    XENO      &            Xe Arc Lamp            & 2 states: On, Off.                                           \\
    NEON      &            Ne Arc Lamp            & 2 states: On, Off.                                           \\
    THAR      &      Thorium-Argon Arc Lamp       & 2 states: On, Off.                                           \\
    DEUT      &     Deuterium Flat Field Lamp     & 2 states: On, Off.                                           \\
     QTH      & Quartz Tungsten Halogen Flat Lamp & 2 states: On, Off.                               \\
\hline
    CROT      &            Co-rotator             & 4 readings: Is moving, Is touched, Is in Fault, Is Active.              \\
    CVTS      &        Cryo-Vacuum sensors        & 13 readings: 10 analog and 3 digital sensor readings from CVS. \\
    IRTS      &         Infrared sensors          & 33 readings: 28 analog and 5 digital sensor readings from NIR. \\
    CPTS      &             CP sensor             & 1 reading: Temperature sensor on common Path.   \\
   THCU1-2    & Thermal Control Unit 1-2 sensors  & 5 readings: analog teperatures and flow rate from instrument cabinets. \\

\hline
\end{tabularx}
\caption{SOXS devices and details. The upper section is related to
  motors, the middle section to lamps, while the lower section is
  related to sensors.}
    \label{tab:devices}
\end{table}

\item[Calibration Unit (CU):] includes two sets of lamps and a linear
  motor responsible for selecting the appropriate calibration source:
  
  \begin{itemize}
  \item
    The ﬁst set is composed of ﬁve arc lamps - Mercury, Argon, Xenon,
    Neon, and Thorium- Argon - used for wavelength calibration (MERC,
    ARGO, XENO, NEON, THAR);
  \item
    the second set is composed of 2 continuum Deuterium and Quartz
    Tungsten Halogen lamps for flat-field calibration (DEUT, QTH);
  \item
    the ``Calibration Unit Pinhole Slide'' linear motor (CUPS) 
    controls an insertable pinhole for alignment and maintenance
    procedure purposes.
  \end{itemize}

\item[Acquisition Camera (ACQ):] centers the source on the
  selected slit during target acquisition. The
  subsystem is composed as follows:
  \begin{itemize}
  \item
    the camera, which can also be used as a light imager to perform
    photometry and flux calibration, is a commercial \emph{Andor
      iKon-M 934 Series} equipped with a deep-depletion CCD and
    providing a $3.5^\prime\times 3.5^\prime$ Field of View (FoV) with
    a resolution of $0.205^{\prime\prime}/\rm px$;
  \item
    ACQ also includes an ``Acquisition Camera Focuser'' linear motor
    (AFOC) for focusing, and an a ``Acquisition Camera Filter Wheel''
    eight positions rotary motor (ACFW) provided with Vimos-$V$,
    Vimos-$Y$, SDSS $ugriz$ filters, and a Free position.

  \end{itemize}

\item[Near-infrared Spectrograph (NIR):] this sub-unit is composed of
  an echelle-dispersed spectrograph working in the $800$ -- $2000 \rm nm$
  wavelength range. It is enclosed in a cryostat supplemented by a
  Cryo-Vacuum controller.  The system comprises:
  
  \begin{itemize}
  \item
    \hl{a \emph{Teledyne H2RG TM} $2048 \times 2048 \rm px$ hybrid
    infrared array detector; its front-end electronics, the custom NGC
    controller developed by ESO, and a mechanical support;}
  \item
    a cryogenic, a piezo-mechanical ``Near-Infrared Slit Exchanger''
    with slits from $0.5$ to $5^{\prime\prime}$, a pinhole and a multiple
    pinhole position for calibration and alignment. 
  \end{itemize}

\item[Visible Spectrograph (UV-VIS):]  based on a novel design in
  which the spectral band is split in narrow sub-bands, thus allowing
  the use of high efficiency gratings that are optimized for a narrow
  wavelength range.  Besides the fixed optics, it comprises:
  
  \begin{itemize}
  \item
    \hl{an \emph{E2V CCD44-82} $2048 \times 4096 \rm px$ CCD detector, its
    front-end electronics and the custom NGC controller developed by
    ESO;}
  \item
    a ``Visible Slit Exchanger'' linear motor (VISE) with the same
    options as the NISE, plus an additional mirror position;
  \item
    the mirror position in the VISE redirects the light source to a
    technical camera (TECH) that was used during the Assembly,
    Integration, and Veriﬁcation (AIV) phase.
  \end{itemize}

\item[Cryo-Vacuum Control Sub-system (CVS):] consists of all
  hardware and the electronics related to controlling and monitoring of
  the cryogenic functions for the two cryostats (enclosing the whole
  NIR spectrograph and the only UV-VIS detector, respectively, as in the
  filled box in Fig.~\ref{fig:opto}).
  
\end{description}

In this paper, we present in detail the SOXS Instrument Control
Software, expanding and structuring the improvements described in a
dedicated series of SPIE proceedings\cite{
  2024SPIE13101E..2GR,
  2020SPIE11452E..2QR,
  2018SPIE10707E..1GR }.

The development and commissioning phase is shown in
Sect.~\ref{sec:dev}. We present the control network architecture
in Sect.~\ref{sec:net} and the overall design of the instrument
software is in Sect.~\ref{sec:sof}.
Sect.~\ref{sec:special} provides further details about the special devices:
\begin{inparaitem}[\null]
\item  the Active Flexure Compensation system (Sect.~\ref{sec:afc});
\item  the co-rotator (Sect.~\ref{sec:crot});
\item  the Near Infrared Slit Exchanger (Sect.~\ref{sec:nise});
\item  the ACQ camera software (Sect.~\ref{sec:cam}).
\end{inparaitem}
A brief description of the Synoptic Panel is given in Sect.~\ref{sec:syn}, and
conclusions are presented in Sect.~\ref{sec:conc}.

\section{Development and Commissioning}
\label{sec:dev}

ESO instruments such as X-Shooter have been developed mainly for VLT
at Paranal Observatory\cite{1997SPIE.3112...20A}, following the VLT
standards and guidelines in terms of hardware and software components.
For what concerns SOXS instrument softwre (INS), the software standard is
the \emph{VLT Common Software} (VLTSW), latest release of which is
VLT2024.

SOXS development was not forced to closely follow the VLT
stantards, being that it is intended for the La Silla
Observatory\cite{2019lsof.confE..21S}.  However, SOXS shares
similarities in design and in many operational procedures to
X-Shooter. For that reason we decided to take the existing
X-Shooter INS as baseline.
Thus, during the design phase we adapted the template instrument
software provided by the VLTSW to the SOXS, and conﬁgured standard
software devices under the VLT2024 release.

Procedures for starting-up and shutting-down the software also follow the
VLT rules.
Automated tests have been implemented to excercise software built from
scratch: testing of the instrument functions, the detectors, and the
software layers and scripts responsible for coordination of all motors,
lamps and detectors to perform observation and maintenance procedures.
Most of the instrument software development was carried out in advance
using software device simulators provided by VLTSW, and the
configuration was tested or adapted gradually with the hardware
becoming available.

After the Assembly, Integration, and Test (AIT) phase in
\emph{INAF -- Osservatorio Astronomico di Padova} \cite{
  2022SPIE12184E..81A,
  2022SPIE12184E..84C,
  2018SPIE10702E..3DB},
the instrument was shipped to La Silla where the AIV phase was
carried out\cite{2024SPIE13099E..1NC} to test the whole
instrument at both engineering and scientific level.
The preliminary operations at La Silla allowed us to validate the
instrument performances with respect to the end-to-end
simulations\cite{
  2024SPIE13099E..05S,
  2022SPIE12187E..0CG,
  2020SPIE11450E..1BG}.

The commissioning phase validated the dedicated scheduler\cite{
  2024SPIE13101E..2FA,
  2022SPIE12189E..0AA
},
preparing the real time the observation queue.
The collected data are processed on-site, a Python reduction pipeline
called \texttt{soxspipe}, which is also publicly available and easy to
install allowing users to set up custom reduction
strategies\cite{
  2022SPIE12189E..0LL,
  2022SPIE12189E..1IY,
  2020SPIE11452E..2DY}.

\section{Network Architecture}
\label{sec:net}

\begin{figure}[t]
  \centering
  \includegraphics[width=\textwidth]{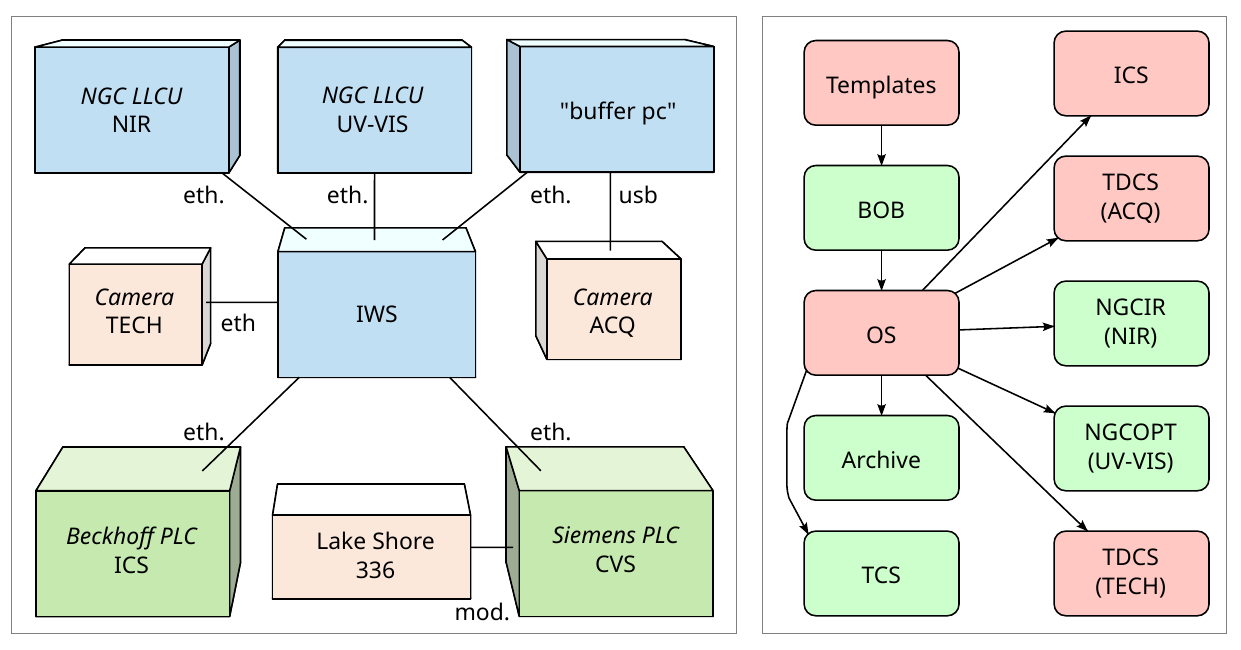}
  \caption[SOXS Network and Software architecture]
  { \label{fig:architecture}
    \textbf{Left}: SOXS network architecture, showing workstations in
    blue, PLCs in green, and other devices in pink. USB, Ethernet and
    Modbus communication protocols are also shown.
    \textbf{Right}: SOXS software architecture. The VLTSW components
    depicted in light green have been re-used without any
    modification.  The components in light-red have been developed by
    the SOXS team, starting from the VLTSW base software provided for
    each type of component. }
\end{figure}

The SOXS network architecture follows the typical configuration of VLT
instrument control systems: through the instrument LAN an Instrument
Workstation (IWS) supervises several connected local controllers,
mostly based on Gb Ethernet (see the left diagram in
Fig.~\ref{fig:architecture}).

The detector controllers for the NIR and UV-VIS channels are 
the ESO New General Detector Controller (NGCs). The respective Linux
Local Control Units (LLCUs) are connected to the IWS via the
instrument LAN.

Following the recently-introduced VLT standard, a single
\emph{Beckhoff} Programmable Logic Controller
(PLC) is responsible for the control of all
instrument functions\cite{
  2024SPIE13096E..2WC,
  2020SPIE11452E..25C,
  2018SPIE10707E..2HC},
i.e., motors, lamps, and most of the SOXS sensors.
A separate \emph{Siemens} S7 PLC is responsible for the CVS functions
and sensors.  The CVS is an autonomous system, SOXS INS is not in
charge of controlling but only monitoring it.
In particular, NIR and UV-VIS temperatures are controlled and
monitored by means of a \emph{Lakeshore} 336 controller connected to
the \emph{Siemens} PLC via modbus protocol.

On the other hand, the ACQ camera is based on a Commercial, Off-The
Shelf (COTS) component, providing an integrated controller with an USB
interface.

\hl{The baseline design routed the USB connection to the IWS through
  the observatory LAN by means of a commercial USB extender. However,
  we found on-site that this led to timeout errors, impacting the
  proper functioning and performance of the system. Therefore, we
  decided to directly USB-attach the camera on a separate fanless PC
  mounted on the SOXS flange, that we call ``buffer pc'', which is
  connected to the SOXS network via the Instrument LAN. This buffer pc
  also runs the VLT2024 release in order to easily set up a ``standard
  VLT way'' for software installation, maintenance, and communication
  between the camera and the IWS.}

Finally, an additional Cameralink Technical Camera (TECH), physically
placed in the UV-VIS spectrograph subsystem, is also linked via Gb
Ethernet.  This camera is not used in normal operations, and is
manually connected to the LAN only for maintenance and diagnostics.

\section{Software Architecture}
\label{sec:sof}



The SOXS INS (see the right diagram in Fig.~\ref{fig:architecture}) is
developed using the latest VLT software release (VLT2024).
The architecture follows the standard partitioning
of VLT Instrument software applications.
It is in charge of controlling all instrument functions and monitoring
all sensors:

\begin{description}
\item[ICS:] \emph{Instrument Control Software} manages all instrument
  functions. This component supports a large number of ``standard
  devices'', like linear stages, shutters, lamps, etc. No code
  development is necessary for these devices and it is enough to
  provide conﬁguration information (location and type of signals,
  motor control parameters, etc.).
  Non-standard devices are required to properly interface with ICS,
  developing a Function Block (FB) software at the PLC level and a
  ``special device'' driver at the IWS level.  These devices are
  detailed in Sect.~\ref{sec:special}.

\item[DCS:] the NIR and UV-VIS spectrograph detectors are controlled
  by instances of the \emph{Detector Control Software}. These are
  ``standard detectors'' and the ICS is required to only provide
  conﬁguration information information and voltages parameters for
  readout modes and detector Regions Of Interest.
  
\item[TDCS:] the ACQ and the TECH cameras are based on instances of
  the \emph{Technical Detector Control Software} Development Kit (SDK)
  \cite{2014SPIE.9152E..0ID}, which has been part of the VLT Common Software
  since the VLT2016 release.
\item[OS:] operations are coordinated by a central component, the
  \emph{Observation Software}, which is also in charge of managing
  external interfaces. All observation, calibration and maintenance
  procedures are implemented as template scripts using the \hl{Tool
    Command Language (TCL) language}, which are executed by the Broker
  of Observation Blocks (BOB) through commands sent to the OS.
\item[TCS:] the \emph{Telescope Control Software} is an external
  interface, running the VLT2010 release.  Its control software has
  been recently adapted to host SOXS as an NTT instrument.
\end{description}
Finally, OS also manages an additional external interface, the
Archive, through another standard component provided by the VLTSW: the
OS Archiver. It is in charge of producing the files to be archived. In
particular, it merges all partial headers, coming from ICS and TCS,
with the header of the image file produced by the DCS. The ICS partial
header contains information about the instrument functions, while the
TCS partial header contains information about pointing, ambient and
observing conditions.  

\section{Special Devices}
\label{sec:special}

In SOXS, special devices requiring custom implementation of the
Function Block at the PLC level and Device Driver at the ICS level are
AFCs, CROT, and NISE.  Details of the development of these special
devices are given in the following sections.

\hl{We have included the Andor camera in this section since it is not a
standard VLT component. Its control software is an application based on
the TDCS and is a special feature of SOXS.}

In the following subsections, we describe the implementation giving
some details about the classes that manage these devices, and their
inheritance. The ones prefixed with \texttt{sx}- are SOXS specific
implementations, while the ones prefixed with \texttt{ic0fb}- are
VLTSW natives.

\subsection{Active Flexure Compensation}
\label{sec:afc}

Since SOXS is installed on \hl{the Nasmyth rotator adapter, which will
  track the sky during an observation, it changes its orientation
with respect to the gravity vector.  This results in some flexures
between the spectrograph and the common path system, which move the
target with respect to the spectrographs slit.}  For this reason, two
piezo-actuated tip-tilt mirrors (AFC1 and AFC2) are located in the CP
and are used to correct for the flexures.

Flexures displace the target in relation to NISE and VISE. To address
this issue, AFCs are commanded by INS through the instrument PLC via
analog signals (one per axis). Since the AFCs are not a VLT standard
actuator, a ``special device'' has been developed.  During
observations, this component operates as a ``tracking axis'', updating
in a closed-loop the position of the devices depending on the rotator
angle.

Since the expected loop frequency is about $1\rm Hz$, timing constraints
are not tight. It has been decided to implement this tracking loop
using a look-up table entirely in the IWS.

The AFCs are controlled by two \texttt{PI S-330} two-axis
actuators. Each actuator is controlled by a \texttt{PI E-727.3SDA} 3
channel digital piezo controller, which is commanded through the
instrument PLC via serial line.  The AFC system operates in the
following modes:

\begin{itemize}
\item Mode AUTO, in which the correction is periodically computed and
  applied (about every minute) by the software on the basis of a
  ``pointing model''. The pointing model requires a calibration
  procedure and the computation of corrections requires information
  about the rotator position. 
\item Mode STAT, in which the AFCs are kept at a fixed position, sent
  via a SETUP command.
\item Mode REF, which puts the AFCs at a fixed, pre-defined, position
  required for the alignment of the system.
\end{itemize}

A dedicated device VLTSW driver class, in this case
\texttt{sxiafcDevDrv}, is derived from \\ \texttt{ic0fbDevDrvBase},
methods are developed to implement the device specific behavior. In
particular, method \texttt{controlLoopUser} encapsulates the logic for
AFCs positioning. The method is called periodically by the underlying
ICS framework code.
If the AFC has been setup with a fixed position (either specified by
the user or the reference one), the (fixed) positioning command is
``refreshed''. If the device must compensate for flexures, a new AFC
command is computed for the current position of the de-rotator and
applied.  The loop period can be set in the device configuration.

In the case of the AFC, commands are sent to the device via serial
line.  On the PLC side we developed a Function Block which uses the
library \texttt{FB\_RS232} provided by ESO, which was recently updated
under our suggestion to take into account the rotator offset.  A
device simulator (\texttt{sxiafcDevSim}) allows the software to
operate in simulation.

\hl{Two custom widgets in the ICS panel allow the operation mode to be
changed and, if STAT is selected, the selection of a custom position.}

\subsection{Co-rotator}
\label{sec:crot}

During operations, the rotation of the whole instrument would also
affect cabling in several devices.
For that reason, a co-rotator system has been set up to compensate for
the de-rotator.
\hl{Moreover, to prevent the risk of damaging the harnesses in case of
mis-alignment between co-rotator and de-rotator, a security system
based on an interlock chain has been set up.}

Two linear potentiometers (Tekkal TR-25) have been mounted on
an interface flange between the instrument and the co-rotator to
check the relative rotation between them.
\hl{These signals, that correspond to the relative displacement
  between the two systems in the rotation, give the feedback to the
  servo motor driver.}
The potentiometers adapter accepts these signals as inputs and returns
their difference as output.  The adapter board output is an input for
a motor driver (Lexium 32C), that controls the speed and the direction
of rotation of the co-rotator motor to match the NTT Nasmyth rotator.  A
safety switch (Omron ZCQ2255) is connected, and is inserted in the
telescope interlock chain in order to disable the rotator when
the co-rotator is in an OFFLINE state.
If the switch is reached, the interlock signal is raised and the final
stage that gives power to the servo motor is disabled. A manual
repositioning of the co-rotator is needed to disable the interlock and
reactivate the final stage.
When the final stage is disabled and the co-rotator is not working,
the ACTIVE output signal is low.  The special device driver is
controlled via digital and analog Input/Output:
\begin{itemize}
\item two analog inputs give information to the driver about the
  relative rotation between the instrument and the co-rotator;
\item a digital input, coming from a safety switch, stops
  the motor in case of emergency / safety risks;
\item two digital input signals, coming from the Beckhoff PLC, 
  control the activation and reset of the driver: ENABLE (Enable the
  power stage); FAULT RESET (Reset error message);
\item two digital output signals  interface with the Beckhoff PLC
  (ES2004 and ES1004 modules)
  giving information about the state of the driver: ACTIVE
  (Indication of the Operating State); NO FAULT (Indication
  of the Fault condition).
\end{itemize}
Again, the custom widget in the ICS panel allows changing the state of
the special device, and to monitor the above flags.

\subsection{Near Infrared Slit Exchanger}
\label{sec:nise}

The NIR Infrared Slit Exchanger (NISE) is a cryogenic actuator
controlled via a Micronix MMC-110 crontroller, connected to
the SOXS PLC through a serial line of type RS232.  Since the
Micronix controller is not directly supported by the VLT Software, a
special device needs to be developed.

A dedicated device driver class named \texttt{sxiniseDevDrv} was
derived from the VLTSW class \texttt{ic0fbDevDrvBase}. Methods were
developed to implement the device specific behavior.
State change handling methods handles setting up of the communication
with the controller.  The setup handling method is overloaded to
transform setup requests into commands for the Micronix controller.
The status handling method is overloaded to retrieve status
information from the Micronix controller, returning it as a command
reply and storing it in the database in order to be displayed in GUIs.

The device server \texttt{sxiniseDevSrv}, i.e. the process that hosts
the driver code, is based on the standard server class
\texttt{ic0fbDevSrv} and makes use of standard communication
interfaces \texttt{ic0fbsiaOpcUa} and \texttt{ic0fbIfCcs} to
communicate with the driver or the simulator.

\subsection{Acquisition Camera Control}
\label{sec:cam}

The imaging camera of SOXS is a COTS component, provided with a USB
2.0 interface for external connection.
Since the SOXS IWS is located in the data center of the La Silla
Observatory, about $3 \rm km$ from the NTT where SOXS and the ACQ are
installed, while USB cable connections are limited to a few meters, we
initially decided to use a commercial, switchable USB extender (Icron
2304GE-LAN) to route the connection through the La Silla LAN and then
to the IWS.  This solution, while it did not present any issue during
the AIT phase in Padova, has not been validated in La Silla. This led
to random, unpredictable timeouts.

Therefore, we switched to a fall-back design where an intermediate
computer, called ``buffer pc'', is installed on the instrument.
Communication from the ``buffer pc'' to the IWS happens through the
Instrument LAN.  This PC is close to the camera and can be connected
directly to it via USB.  Provided with the VLT2024 software release
and specific SOXS modules, the ``buffer pc'' runs the ACQ and is
responsible for passing the information towards the IWS.

The ACQ camera control is based on TDCS\cite{2014SPIE.9152E..0ID} SDK,
and its core consists of a class providing the ``communication
interface'' with the camera.  This class directly interfaces with the
Andor camera via USB, directly calling the vendor-supplied driver
functions.

\section{Synoptic Panel}
\label{sec:syn}

\begin{figure}[t]
  \centering
  \includegraphics[width=\textwidth]{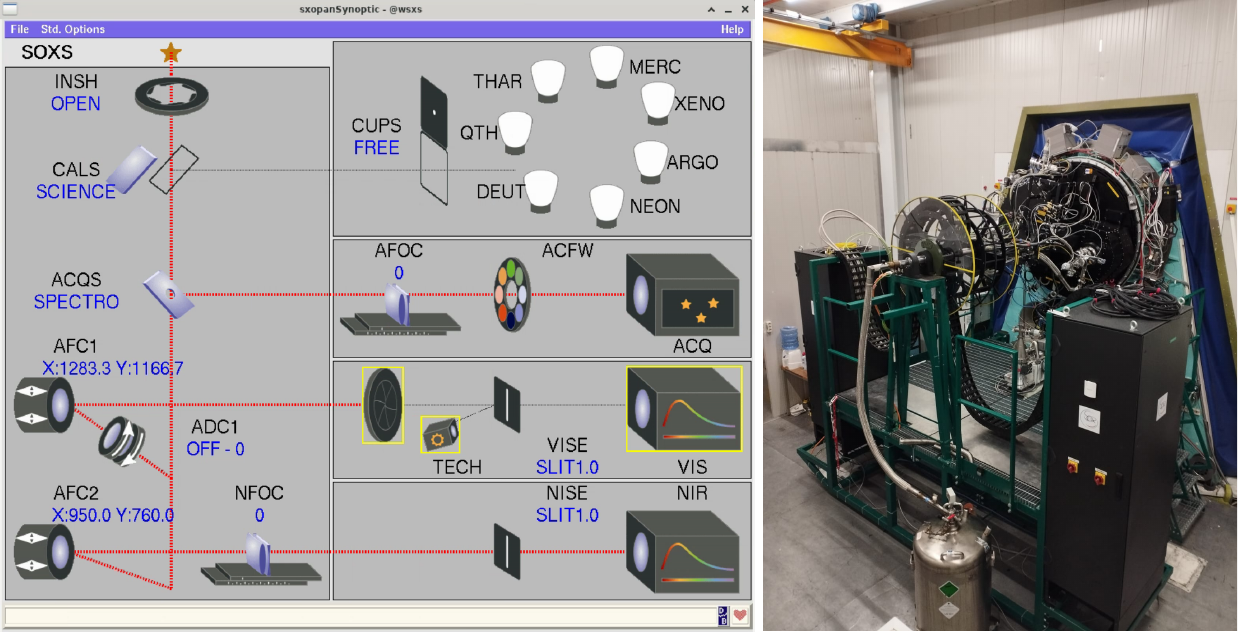}
  \caption[SOXS Synoptic Panel]
  { \label{fig:synoptic}
    \textbf{Left}: SOXS Synoptic Panel. Devices names are shown in
    black; while corresponding state or positions are shown in blue.
    In this image, yellow borders indicate that a component is
    simulated.  The red path indicates the light flow into the
    instrument. \textbf{Right}: A picture of the SOXS instrument
    installed at NTT.}
\end{figure}

The Synoptic Panel (see left panel in Fig.~\ref{fig:synoptic}) is a
custom developed GUI showing the current configuration of the
instrument in a graphical way.

This panel displays the light path, the current position/status of the
motors and lamps, and if any of these components are in simulation
mode.

ESO recently requested the Synoptic Panel for recent instruments, such
as SPHERE, in order to have a quick look at their status.  SOXS is the
first La Silla instrument to provide it, \hl{and we have received
  positive feedback from both ESO/La Silla operators and SOXS users.}

In general, this kind of panel follows the opto-mechanical design of
the instrument, where components and the light path are reproduced to mock
their ``real'' position (see a picture of the whole instrument in
Fig.~\ref{fig:synoptic} on the right).
For SOXS, we decided to change the approach towards a ``subway map''
concept, to improve the operator's readability.

In particular, most of the gif icons animate depending on their status
(shutter OPEN or CLOSE, calibration unit lamps ON or OFF, type of the
slits, position of the mirrors), and the light path is set to suggest
where the light is ``flowing'' into the instrument.  This is
particularly useful in the case of the UV-VIS shutter, which is
autonomously controlled by the corresponding NGC.
Finally, simulated devices are bordered to improve awareness.
Given the positive feedback by the team, we decided to broaden this
approach also for the ESO VLT ERIS instrument and for the LBT
SHARK-NIR coronagraph\cite{2025JATIS..11b5005R}. Feedback also
suggested the panel to be interactive, i.e. allow click-and-setup of
single components instead of bare monitoring.  This intriguing option,
which is already available in SOXS ``traditional'' engineering panels
developed for each subsystem, will be explored in the framework of
the next telescope instruments.

\section{Conclusion}
\label{sec:conc}

We presented the general software architecture of the SOXS instrument
for the NTT, a double spectrograph for near-infrared and visible
observations. For the Acquisition Camera, that is available for
``light imaging'', we described the peculiar features in terms of
hardware (Acquisition Camera and camera connection) and software
solutions (special devices).

The instrument software is based on the VLT Common Software for the
management of standard devices, and on custom software for the
management of special devices, i.e. the piezo-electric actuators for
flexure compensation, the co-rotator device, the piezo-mechanic slit
exchanger for the near-infrared spectrograph, and the Acqusition
Camera.

Spectrograph detectors are managed using ESO NGC controllers, while
the COTS imaging camera is managed through a custom software based on
the VLTSW TDCS component.

Other peculiarities in the development of the SOXS Instrument Control
Software include the Synoptic panel for the quick look of the
instrument status and setup.
\hl{During the installation, an on-site major change in the design was
necessary to overcome timeout errors due to the USB extender initially
used fot the Acquisition Camera communication.}

Observation and Maintenance software have been completed and SOXS is
successfully in operation.

\section*{Disclosures}

The authors declare that there are no financial interests, commercial
affiliations, or other potential conflicts of interest that could have
influenced the objectivity of this research or the writing of this
paper.

\section*{Code and Data}

Data sharing is not applicable to this article.
There is no software code associated with the main claims of this
manuscript.  This paper does not rely on supporting data.

\acknowledgments

The authors thank INAF for funding the project. The authors also
acknowledge the support of FONDECYT Postdoctoral Project No. 3180673.
  
\section*{Biography}

The first author is MSc in Astronomy and PhD in Sciences in the field
of Astrophysics.  Currently technologist at the INAF - Osservatorio
Astronomico di Padova, he is involved in Instrument Control Software
of ESO and LBT projects, as well as in the development of small
observatories for science and outreach purposes.  His interests span
from web user interfaces for telescopes, to extrasolar planets
observations.

\listoffigures


\bibliographystyle{spiejour}   
\bibliography{biblio} 


\end{document}